\documentclass[floats,floatfix,showpacs,amssymb,prd,twocolumn,superscriptaddress,nofootinbib,nolongbibliography,reprint]{revtex4-2}

\usepackage{amssymb,amsmath,verbatim,mathtools,needspace,enumitem,etoolbox,graphicx,physics,microtype,afterpage,bigints,gensymb,tabularx,xspace,bm}%
\usepackage{multirow}

\usepackage{xfrac}
\usepackage[dvipsnames, usenames]{xcolor}
\definecolor{linkcolor}{rgb}{0.0,0.3,0.5}
\definecolor{dodgerblue}{HTML}{1E90FF}
\usepackage[T1]{fontenc}
\usepackage[utf8]{inputenc}
\usepackage{orcidlink}
\usepackage{tensor}
\usepackage{graphicx} %

\makeatletter
\newcommand*{\balancecolsandclearpage}{\close@column@grid \cleardoublepage \twocolumngrid}
\makeatother

\def\apj{\rm{ApJ}}
\def\apjl{\rm{ApJ}}

\def\aap{\rm{A\&A}}

\def\mnras{\rm{MNRAS}}

\def\icarus{\rm\textit{Icarus}}

\newcommand{\columbia}{\affiliation{Columbia Astrophysics Laboratory, Columbia University, 550 West 120th Street, New York, NY 10027, USA}}

\newcommand{\oldedits}[1]{{\leavevmode{#1}}}
\newcommand{\new}[1]{{\leavevmode{#1}}}

\begin{document}

\title{Inferring Eccentricity of Binary Black Holes from Spin-Orbit Misalignment}%

\author{Vishal Baibhav$\,$\orcidlink{0000-0002-2536-7752}}
\thanks{NASA Einstein Fellow}\email{vb2630@columbia.edu}
\columbia


\date{\today}

\begin{abstract}

Orbital eccentricity remains one of the least accessible parameters in observations of binary black hole (BBH) systems, largely erased by gravitational radiation long before detection. We introduce a new method to recover this lost parameter by using a more accessible and routinely measurable quantity: spin-orbit misalignment.  In isolated binary evolution, a natal kick from the second supernova both tilts the orbital plane and injects orbital eccentricity, forging a direct and quantifiable connection between spin-tilt and post-supernova eccentricity. By measuring this spin-tilt using gravitational waves, we can not only constrain the natal kick, but we can also reconstruct the binary's formation eccentricity.
We apply this method to GW190412 \new{and GW241011}, assuming an isolated formation channel, and show how their eccentricity at formation can be constrained even in the absence of direct eccentricity measurements. As more advanced detectors come online, improved signal-to-noise ratios will tighten spin-tilt constraints, allowing more precise and reliable estimates of BBH formation eccentricity. \oldedits{Combining this method with multiband observations from LISA and next-generation (XG) detectors will allow us to recover not only eccentricity but also the binary’s orbital separation and redshift at formation, offering a clearer picture of the birth environments of BBH systems and processes that drive their merger.}

\end{abstract}

\maketitle

\section{Introduction}

Gravitational wave (GW) astronomy has transformed our understanding of black hole binaries with routine measurements of their masses, spins and redshifts. However, orbital eccentricity remains one of the final frontiers in fully characterizing these systems.
As a powerful discriminant between formation channels, it has become a topic of intense interest, motivating new developments in waveform modeling~\cite{Damour:2004bz,Memmesheimer:2004cv,Konigsdorffer:2006zt,Buonanno:2006ui,Husa:2007rh,Hinder:2008kv,Arun:2009mc,Yunes:2009yz,Mroue:2010re,Purrer:2012wy,Huerta:2014eca,Moore:2016qxz,Tanay:2016zog,Huerta:2016rwp,Huerta:2017kez,Cao:2017ndf,Healy:2017zqj,Hinder:2017sxy,Hinderer:2017jcs,Ramos-Buades:2018azo,Ramos-Buades:2019uvh,Tiwari:2019jtz,Moore:2019xkm,Pan:2019anf,Phukon:2019gfh,Chen:2020lzc,Wette:2020air,Chiaramello:2020ehz,Ramos-Buades:2021adz,Liu:2021pkr,Islam:2021mha,Setyawati:2021gom,Riemenschneider:2021ppj,Khalil:2021txt,Nagar:2021gss,Cho:2021oai,Paul:2022xfy,Xu:2022zza,Knee:2022hth,Ramos-Buades:2022lgf,Albanesi:2022xge,Joshi:2022ocr,Albertini:2023aol,Liu:2023ldr,Carullo:2023kvj,Henry:2023tka,Shaikh:2023ypz,Albanesi:2023bgi,Ficarra:2024nro,Nagar:2024dzj,Islam:2024bza,Islam:2024zqo,Islam:2024vro,Islam:2024rhm,Islam:2024tcs,Bonino:2024xrv,Boschini:2024scu,Gamboa:2024hli,Gamboa:2024imd,Paul:2024ujx,Fumagalli:2025rhc,Islam:2025llx,Islam:2025rjl,Manna:2024ycx,Shi:2024age}, data analysis searches~\cite{Tiwari:2015gal,Gondan:2017hbp,Lower:2018seu,Brown:2009ng,LIGOScientific:2019dag,Romero-Shaw:2019itr,Nitz:2019spj,Wu:2020zwr,Romero-Shaw:2020thy,Lenon:2020oza,Ramos-Buades:2020eju,Gayathri:2020coq,OShea:2021faf,Romero-Shaw:2021ual,Gamba:2021gap,Cheeseboro:2021rey,Wang:2021qsu,Favata:2021vhw,Romero-Shaw:2022xko,Iglesias:2022xfc,Romero-Shaw:2022fbf,Guo:2022ehk,Dhurkunde:2023qoe,Divyajyoti:2023rht,Ravichandran:2023qma,LIGOScientific:2023lpe,Saini:2023wdk,Ramos-Buades:2023yhy,Fumagalli:2023hde,Patterson:2024vbo,Gupte:2024jfe,Planas:2025jny,Morras:2025xfu}, and astrophysical theory~\cite{OLeary:2008myb,Kocsis:2011jy,Antonini:2012ad,Antognini:2013lpa,Silsbee:2016djf,Antonini:2017ash,Rodriguez:2017pec,Samsing:2017rat,Samsing:2017xmd,Samsing:2017oij,Gondan:2017wzd,Rodriguez:2018jqu,Rodriguez:2018pss,Takatsy:2018euo,Liu:2018vzk,Zevin:2018kzq,Fragione:2019dtr,Fragione:2019hqt,Rasskazov:2019gjw,Kremer:2019iul,Liu:2019gdc,Michaely:2019aet,Antonini:2019ulv,Gondan:2020svr,Michaely:2020ogo,Tagawa:2020jnc,Grobner:2020drr,Samsing:2020tda,Zevin:2021rtf,Kritos:2022ggc,Fumagalli:2024gko,Stegmann:2025shr}.
The biggest challenge in measuring eccentricity is that it is one of the first parameters to be erased by GW emission~\cite{Peters:1963ux,
Hinder:2007qu}. Any initial eccentricity a binary possesses is rapidly radiated away during its long inspiral, leaving most systems nearly circular by the time they enter the detection band of current instruments. This makes the direct measurement of eccentricity very difficult.

This article presents an alternative pathway. Instead of attempting a direct measurement, we propose a method to infer the eccentricity of a BBH system immediately following its formation through isolated field evolution. 
The main idea is to use spin-orbit misalignment as a measurable proxy for the eccentricity introduced by the second supernova. During the core collapse of a BH, mass loss and natal kicks reshape the binary's orbit. The natal kick, a consequence of asymmetric mass ejection or neutrino emission, can tilt the orbital plane~\cite{Kalogera:1999tq}. Simultaneously, the combination of the kick and the mass loss alters the orbit's size and shape, inducing a final eccentricity, $e_f$~\cite{1983ApJ...267..322H,1995MNRAS.274..461B}.
In this work, we connect the orbital eccentricity following the core collapse of the second star to a measurable GW signature: the spin-orbit misalignment.
 We apply this framework to the binary black hole systems GW190412 \new{and GW241011}, assuming they formed through isolated binary evolution, and demonstrate how their formation eccentricity can be inferred. We also explore how future detector improvements will enhance the precision of such measurements \oldedits{and how multiband observations will allow us to accurately measure the binary’s orbital separation and redshift at formation.}

\section{The link between spin tilt and eccentricity}

\begin{figure*}[htbp]
    \centering
    \includegraphics[width=\textwidth]{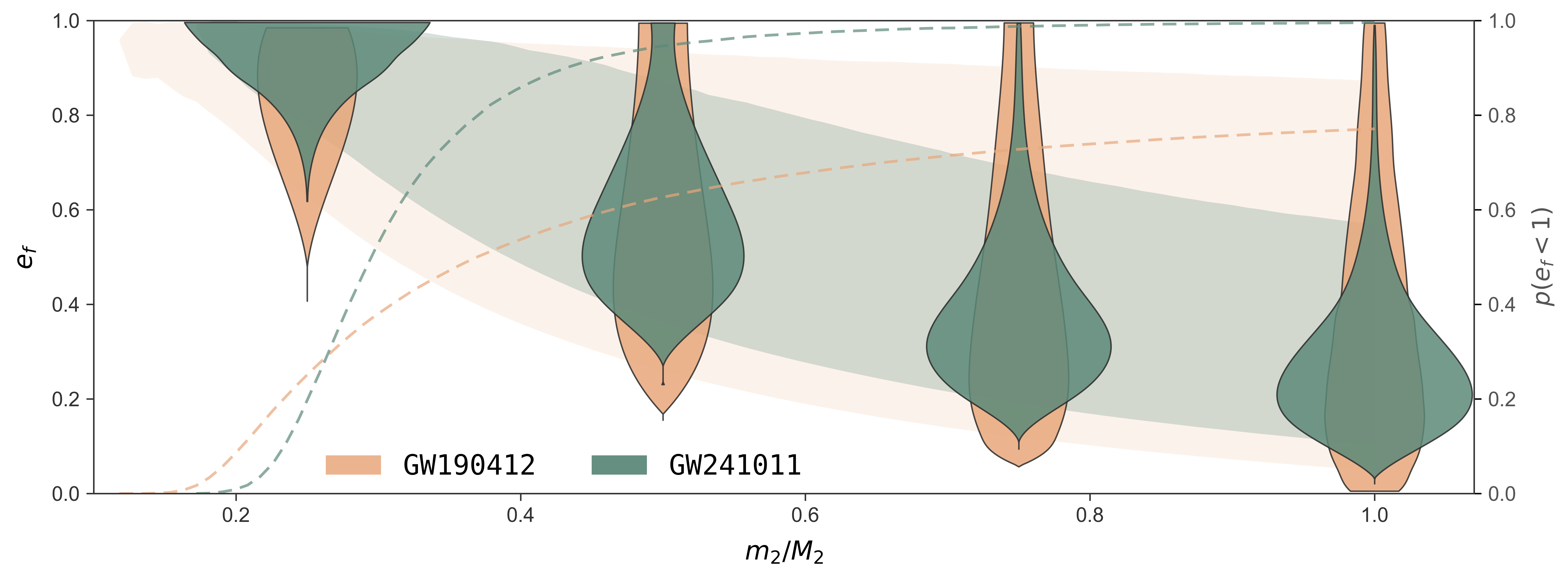}
    \caption{Inferred posterior distribution of the formation eccentricity ($e_f$) of GW190412\new{ (orange) and GW241011 (green)}, shown as a function of the $m_2/M_2$ (representing mass loss during the second supernova). The violin plots show the probability density of $e_f$ for $m_2/M_2 = 0.25$, $0.5$, $0.75$, and $1.0$ (no mass loss). \new{The continuous shaded regions represent the $90\%$ confidence intervals for $e_f$ across all $m_2/M_2$}. The dashed lines, corresponding to the right y-axis, plot the survival probability, $p(e_f < 1)$.
}
    \label{fig:violin}
\end{figure*}

\begin{figure*}[t]
    \centering
    \includegraphics[width=\linewidth]{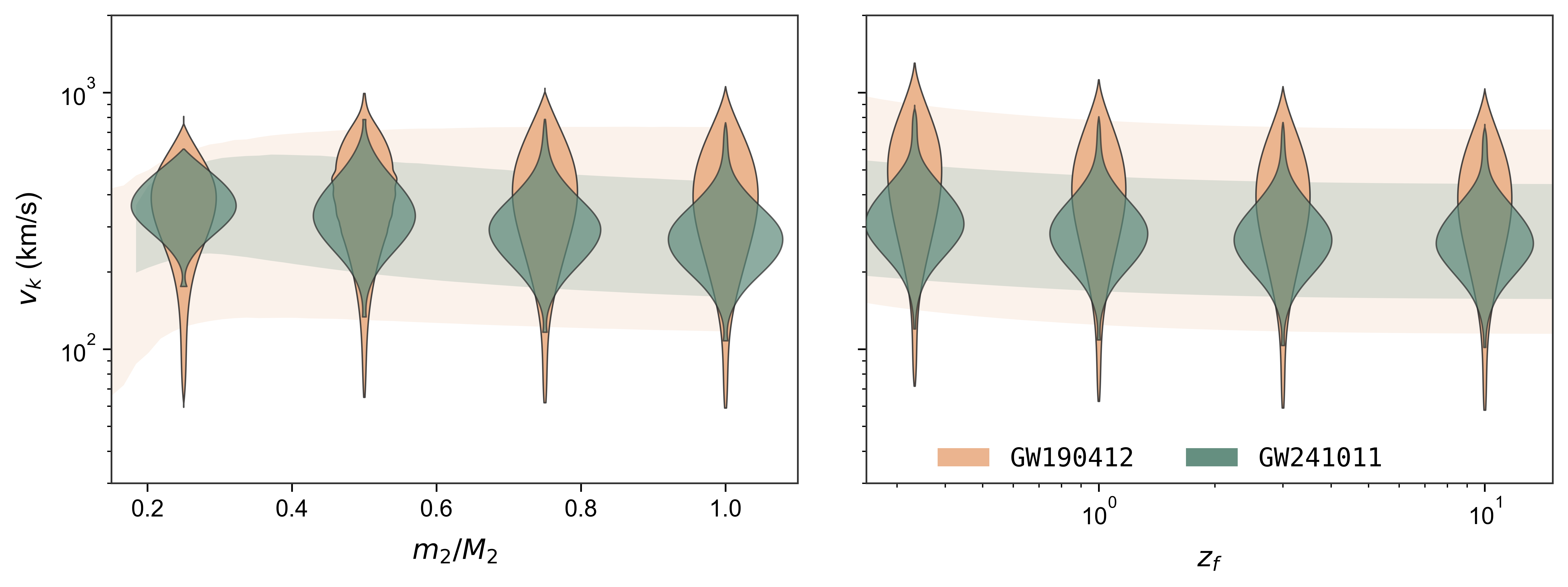}
    \caption{Posterior distribution of the inferred natal kick velocity ($v_k$) imparted to the secondary black hole in GW190412\new{ (orange) and GW241011 (green)}. Left panel: $v_k$ as a function of the fractional mass ratio $m_2/M_2$, assuming a fixed formation redshift of $z_f = 3$. Violin plots show the posterior density of $v_k$ at representative values of $m_2/M_2 = 0.25$, $0.5$, $0.75$, and $1$. Right panel: $v_k$ as a function of the formation redshift $z_f$, assuming $m_2 = M_2$. Violin plots show the posterior density at $z_f = 1/3$, $1$, $3$, and $10$. \new{The continuous shaded regions represent $90\%$ confidence intervals for $v_k$ for full range of $m_2/M_2$ and $z_f$}. 
}
    \label{fig:violin_kick}
\end{figure*}

We build our model on a set of standard assumptions in isolated binary evolution. First, we consider a binary system that has already experienced one supernova, forming the first black hole with mass $m_1$. During a subsequent mass transfer phase, the black hole’s spin is expected to become aligned with the orbital angular momentum~\cite{1996MNRAS.282..291S,1998ApJ...506L..97N,2005MNRAS.363...49K,Bogdanovic:2007hp,Martin:2007jf,Perego:2009cw,ColemanMiller:2013jrk,Banerjee:2019ieq}. In addition, tidal forces act to circularize the orbit~\cite{1966Icar....5..375G,1977A&A....57..383Z,1981A&A....99..126H,1995A&A...296..709V,1995A&A...296..709V,1998ApJ...499..853E,
Hurley:2002rf}, effectively reducing the initial orbital eccentricity to zero. We focus on the moment of the second supernova, where the progenitor star of mass $M_2$ collapses to form the second black hole of mass $m_2$ and finalizes the BBH system. This event is characterized by a sudden mass loss to the star and the impartation of a natal kick to the newly formed BH. Based on observations of young pulsars~\cite{Kramer:2003au, Johnston:2005ka, Wang:2005jg, Ng:2006vh, Johnston:2007gx, Chatterjee:2009ac, Noutsos:2012dt, 2013MNRAS.430.2281N, 2021NatAs...5..788Y}, we assume that the natal kick is aligned with the spin of the newly formed neutron star or black hole and that this spin is aligned with the orbital angular momentum of the progenitor binary (see Ref.~\cite{Baibhav:2024rkn} for caveats). Following the second core collapse, the imparted natal kick, $\vec{v}_k$, adds vectorially to the orbital velocity of the progenitor star, $\vec{v}_{\text{orb}}$, to produce the new orbital velocity of the compact object. This change in velocity alters the direction of the orbital angular momentum vector, leading to a misalignment between the (pre-existing and fixed) spin axis of the first-born BH and the (new) orbital angular momentum of the final BBH system.  The angle of this change, which corresponds to the measurable misalignment between the first BH's spin and the final orbital plane, is denoted by $\theta$. One can derive a simple geometric relationship between the kick velocity and the tilt angle~\cite{Kalogera:1999tq}:

\begin{equation}\label{eq:vk}
\frac{v_k}{v_{\text{orb}}} = \tan\theta\, .
\end{equation}

The final eccentricity, $e_f$, depends on the ${v_k}/{v_{\text{orb}}}$ (which can be inferred from the spin tilt) and the change in the total mass of the system. It can be expressed in terms of the spin-tilt $\theta$ as~\cite{1995MNRAS.274..461B, Kalogera:1999tq}:

\begin{equation}\label{eq:ef}
e_f = \frac{1}{ \cos^2\theta} \cdot \frac{m_1 + M_2}{m_1 + m_2} - 1 \, .
\end{equation}

This equation is the cornerstone of our proposed method: a measurement of the spin-tilt angle $\theta$ from gravitational waves can be translated into a measurement of the eccentricity at formation. \oldedits{Mass loss raises eccentricity, with the no-mass-loss case giving the minimum. If $M_2$ is unknown, a model-independent lower bound is $e_{f,\rm min} = \tan^2\theta$, setting a hard floor based solely on the measured spin-orbit tilt.

For a bound binary ($e_f < 1$), the tilt is limited by $\cos\theta > \sqrt{\frac{m_1+M_2}{2(m_1+m_2)}}$; zero mass loss allows $\theta < 45^\circ$, with increasing mass loss reducing the maximum survivable tilt.
}

\oldedits{The relations presented above use} standard assumptions grounded in physical arguments or observational evidence: complete orbital circularization from tides, a natal kick imparted along the progenitor's spin, and spin-orbit alignment due to tides or accretion. While these assumptions simplify the model, they can be relaxed in more comprehensive studies. 
Nevertheless, the core physical link remains: the same supernova kick that tilts the orbit also induces eccentricity. Thus, the fundamental principle of using spin-orbit misalignment as a tracer for formation eccentricity still holds true, even if the precise relationship is modified.

\oldedits{This method to infer formation eccentricity has important implications for addressing current challenges in GW astronomy. Eccentricity and spins are coupled during binary inspiral~\cite{Fumagalli:2023hde}. 
Even undetectably small eccentricities can bias inferred spin distributions at black hole formation~\cite{Fumagalli:2024gko}. Knowing the formation eccentricity, as proposed for isolated binaries, could help correct this bias.
This approach also enables future multiband observations, where complementary measurements of eccentricity at formation and detection can be used to fully characterize a binary’s formation environment.}

\section{Case Studies: GW190412 \& GW241011}
\new{The GW events GW190412~\cite{LIGOScientific:2020stg} and GW241011~\cite{LIGOScientific:2025brd} serve as useful benchmarks for our framework.
GW190412 resulted from the merger of black holes with component masses of $34.26_{-4.7}^{+5.72},M_\odot$ and $9.63_{-1.09}^{+1.27},M_\odot$, featuring a well-measured primary spin of $0.36_{-0.16}^{+0.14}$ and a spin-orbit misalignment of $\theta = 34.3^{+20.6}_{-26.8}{}^\circ$.
Similarly, GW241011 involved component masses of $m_1 = 18.62_{-2.08}^{+2.06},M_\odot$ and $m_2 = 6.18_{-0.53}^{+0.68},M_\odot$, with an inferred spin-orbit tilt of $\theta = 26.77_{-8.90}^{+10.60}{}^\circ$\footnote{These tilt estimates are based on the IMRPhenomXPHM waveform model~\cite{gwtc2p1_deglitched_data} for GW190412 and IMRPhenomXO4a for GW241011~\cite{LIGOScientific:2025brd}. For both systems, other waveform models yield somewhat larger tilt estimates, but we adopt the more conservative IMRPhenom-based results to ensure a higher likelihood of a bound binary.}.  
While these spin misalignments could be explained by dynamical formation in high-metallicity environments with very large escape velocities~\cite{Baibhav:2020xdf,Hamers:2020huo,Gerosa:2020bjb,Liu:2020gif,Rodriguez:2020viw,Safarzadeh:2020qrc}, they can also arise from large natal kicks in isolated binary evolution. In this work, we proceed under the assumption that both systems formed through isolated binary evolution.
} Given its larger mass, it is reasonable to assume the primary BH formed first.

\oldedits{
To translate the observed spin-orbit misalignments into a constraint on its formation eccentricity, we use Eq.~(\ref{eq:ef}). Since the progenitor mass $M_2$ is not directly measurable, we parametrize the degree of mass loss in the second supernova by the ratio $m_2/M_2$. For each assumed value of $m_2/M_2$, we propagate the posterior distribution of $m_1$, $m_2$ and $\theta$  through Eq.~(\ref{eq:ef}) to obtain corresponding posteriors for the formation eccentricity $e_f$. The results are shown in Fig.~\ref{fig:violin}: violin plots illustrate representative eccentricity distributions for $m_2/M_2 = \{0.25, 0.5, 0.75, 1\}$,} \new{while the shaded regions summarize the confidence intervals over the full continuous parameter space.}

\new{Several conclusions emerge from this exercise, particularly when comparing the constraints derived from both GW190412 and GW241011.

First, the precision of the inferred formation eccentricity is fundamentally limited by the uncertainty in the spin–tilt angle ($\theta$). For GW190412, the broad posterior on $\theta$ allows both nearly circular ($e_f \approx 0$) and highly eccentric ($e_f \to 1$) configurations, depending on the assumed mass loss. Under the no–mass-loss assumption ($m_2 = M_2$), the inferred eccentricity is $e_f = 0.34_{-0.29}^{+0.53}$. In contrast, the tighter spin–tilt constraint for GW241011 yields a better-determined eccentricity of $e_f = 0.25_{-0.15}^{+0.32}$ under the same assumption.

Second, binary survivability provides an additional constraint. The dashed lines in Fig.~\ref{fig:violin} show the survival probability, $p(e_f < 1)$. With no mass loss, GW241011 has a near-certain survival probability of $99.6\%$, compared to $\sim77\%$ for GW190412. For both systems, survival drops sharply as mass loss increases; when $m_2/M_2 \lesssim 0.1$, no posterior supports a bound binary.

The key takeaway is that spin-orbit misalignment can be mapped into meaningful constraints on binary formation eccentricity.  Events such as GW241011, which provides a notably tight constraint on the tilt angle $\theta$, show that this indirect approach is already a powerful tool for inferring the formation eccentricity of BBHs. These case studies demonstrate the method’s potential and motivate its use in future work as spin–tilt measurements become more precise.}

We can also estimate the magnitude of the natal kick, $v_k$, imparted to the second black hole using Eq.~(\ref{eq:vk}). While the spin tilt angle allows us to determine the ratio of the kick velocity to the orbital velocity ($v_k / v_{\text{orb}}$) directly via Eq.~(\ref{eq:vk}), extracting the absolute kick magnitude $v_k$ requires knowledge of the pre-supernova orbital velocity, $v_{\text{orb}}$. This, in turn, depends on the binary’s orbital separation prior to the second black hole's formation, a quantity not directly observable.

By assuming a formation redshift $z_f$, we can estimate the total inspiral time (or delay time) between binary formation and merger. This delay time depends on the post-supernova orbital separation $a_f$ and eccentricity $e_f$, as well as the component masses, which are already well constrained. Since the spin tilt yields an estimate of $e_f$, we can infer $a_f$ from the delay time. The kick velocity can then be written as:
\begin{equation}
v_k = \tan\theta \sqrt{\frac{G(m_1 + M_2)}{a_f(1 - e_f)}}
\end{equation}
Here, $(1 - e_f)$ in the denominator accounts for the orbital expansion due to both the natal kick and any mass lost during the supernova.

Our estimate of the kick magnitude $v_k$ thus depends primarily on two astrophysical uncertainties: the fraction of mass lost during the second black hole’s formation (parameterized by $m_2 / M_2$), and the formation redshift $z_f$. In Figure~\ref{fig:violin_kick}, we show the inferred posterior distribution for the natal kick velocity imparted to the secondary black hole in \new{GW190412 (orange) and GW241011 (green)}. Across a range of $m_2/M_2$ and $z_f$, we find that typical kick magnitudes lie in the hundreds of km/s. For example, under a conservative assumption of no mass loss ($m_2 = M_2$) and a formation redshift near the peak of cosmic star formation (``cosmic noon,'' $z_f=3$), \new{we infer $90\%$ credible intervals of $117$--$734\ \mathrm{km\,s^{-1}}$ for GW190412 and $160$--$449\ \mathrm{km\,s^{-1}}$ for GW241011.} While higher than typically expected, the inferred kick magnitudes are comparable with velocities measured for some black holes in X-ray binaries~\cite{Mirabel:2001ay,Fragos:2008hg,Repetto:2015kra,Repetto:2017gry,Atri:2019fbx,Brown:2023roz} and with predictions from some core-collapse supernova simulations~\cite{Burrows:2023ffl,Burrows:2023nlq,Burrows:2024pur,Burrows:2024wqv}.
The inferred kick magnitude shows only a weak dependence on the assumed formation redshift. This is because $v_k \propto t_{\text{delay}}^{-1/8}$, making the result relatively insensitive to even large variations in delay time.

While we treat $z_f$ as a free parameter here, future work could incorporate astrophysically motivated priors on formation redshift, informed by the cosmic star formation history or a theoretical delay-time distribution~\cite{Fishbach:2021mhp,Fishbach:2023pqs}. \oldedits{Similarly, the prior on mass loss during supernovae could be guided by prescriptions from core-collapse simulations~\cite{Fryer:1999ht,2012ApJ...749...91F,Spera:2017fyx,2015MNRAS.451.4086S,Ertl:2015rga,Sukhbold:2015wba,Mandel:2020cig,Mandel:2020qwb,Renzo:2022rnk}.}

\section{Future Prospects with Advanced Detectors}

\begin{figure}[tbp]
    \centering
    \includegraphics[width=\columnwidth]{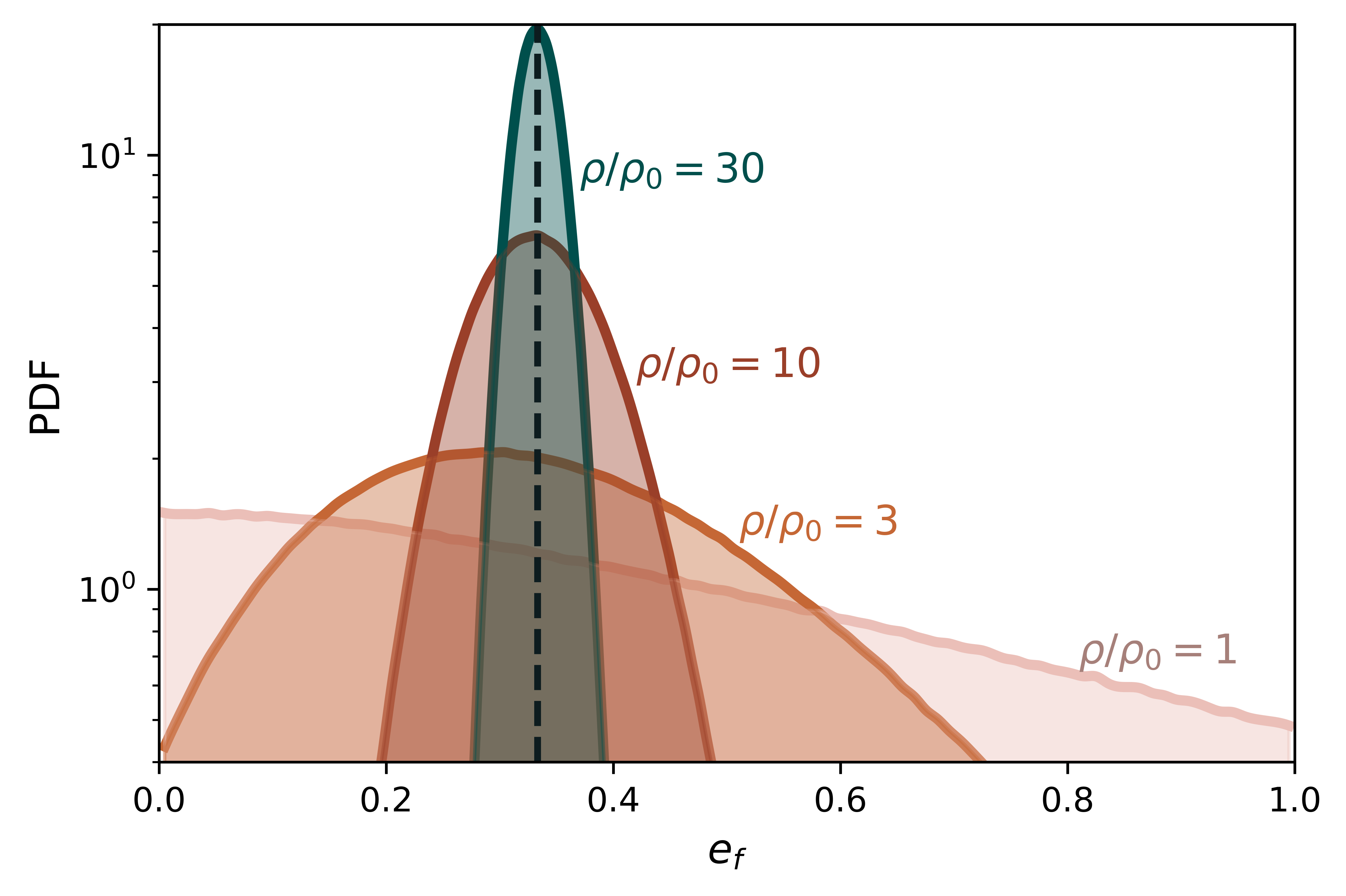}
    \caption{Projected constraints on the BBH formation eccentricity ($e_f$) as a function of improved detector sensitivity. We model a GW190412-like binary with a spin-tilt of $30^\circ$ and assumes no mass loss, for which the theoretical eccentricity is $e_f = 1/3$ (vertical dashed line). Each curve represents the posterior probability density function (PDF) for $e_f$ at a different signal-to-noise ratio (SNR), indicated by the SNR gain $\rho/\rho_0$ \new{with respect to GW190412}. As the SNR increases, the constraint on the spin-tilt tightens, resulting in a progressively narrower and more sharply peaked PDF for the eccentricity.
}
    \label{fig:future}
\end{figure}

The precision of our eccentricity inference is fundamentally limited by the measurement uncertainties in the spin-tilt angle, $\theta$. However, the landscape of gravitational-wave astronomy is rapidly evolving. As current detectors undergo significant upgrades (such as to A+) and next-generation observatories like the Einstein Telescope and Cosmic Explorer are developed, the signal-to-noise ratio (SNR) for events will increase dramatically. This enhancement in sensitivity will directly translate to a more precise measurement of the parameters of binary systems, including spin~\cite{Vitale:2016icu}. The propagated error on the eccentricity, assuming no mass loss for simplicity, is given by:

\begin{equation}
    \Delta e_{f} = \frac{2 \sin\theta}{\cos^3\theta} \Delta\theta < 4 \Delta\theta
\end{equation}
 If future detectors can measure a spin-tilt with a precision of just $1^\circ$, the minimum BBH formation eccentricity could be constrained to better than $0.07$. For a system like GW190412, with a tilt near $30^\circ$, this would correspond to a $1\sigma$ error of $0.027$.

%
%

To illustrate improvement in measuring $e_f$, let's consider a binary system similar to GW190412, with a tilt measurement of $\theta=30^\circ$ (yielding $e_f=1/3$ for no mass loss scenario) and an associated \oldedits{$1 \sigma$} tilt uncertainty of approximately $15^\circ$ at an SNR with current detectors, which we'll denote as $\rho_0=19$. We can project the improvement in our eccentricity measurement by modeling the shrinking error on the tilt as $\Delta\theta = 15^\circ/(\rho / \rho_0)$ where $\rho$ is the new, higher SNR achieved by an advanced detector. Assuming the posterior for $\cos\theta$ is a truncated Gaussian with the corresponding mean and standard deviation, we can generate the resulting posteriors for eccentricity.

The prospects for precisely measuring BBH formation eccentricity will improve dramatically with upcoming detector upgrades and next-generation observatories.
As shown in Fig.~\ref{fig:future}, with increasing SNR, the posterior probability distribution for the BBH formation eccentricity, $e_f$, would become progressively narrower and more sharply peaked around its true value ($e_f = 1/3$ in this example).   In the near term, A+ era detectors are expected to observe a few events per year with a threefold increase in SNR compared to GW190412. For these events, the uncertainty on the  BBH formation eccentricity ($\Delta e_f$) could shrink enough to confidently distinguish moderately eccentric systems from those with zero or near-maximal eccentricity. Furthermore, A+ could detect rare, exceptionally loud events--perhaps one per year--with a tenfold increase in SNR, providing an exceptionally precise measurement $\Delta e_f \approx 0.06$. \oldedits{Next-generation (XG)} detectors are projected to detect hundreds of similar events annually at a $3$x SNR improvement and tens of events per year with an SNR increase of $10$ or more~\cite{Baibhav:2019gxm,2019CQGra..36v5002H,Borhanian:2022czq,Gupta:2023lga}. For these high-SNR events, the uncertainty on eccentricity is expected to reach the percent level ($\Delta e_f \sim {\mathcal O}(0.01)$), turning this indirect inference into a high-precision tool for systematically studying the outcomes of supernova dynamics across a large population of binary black holes.

\begin{figure}[tbp]
    \centering
    \includegraphics[width=\columnwidth]{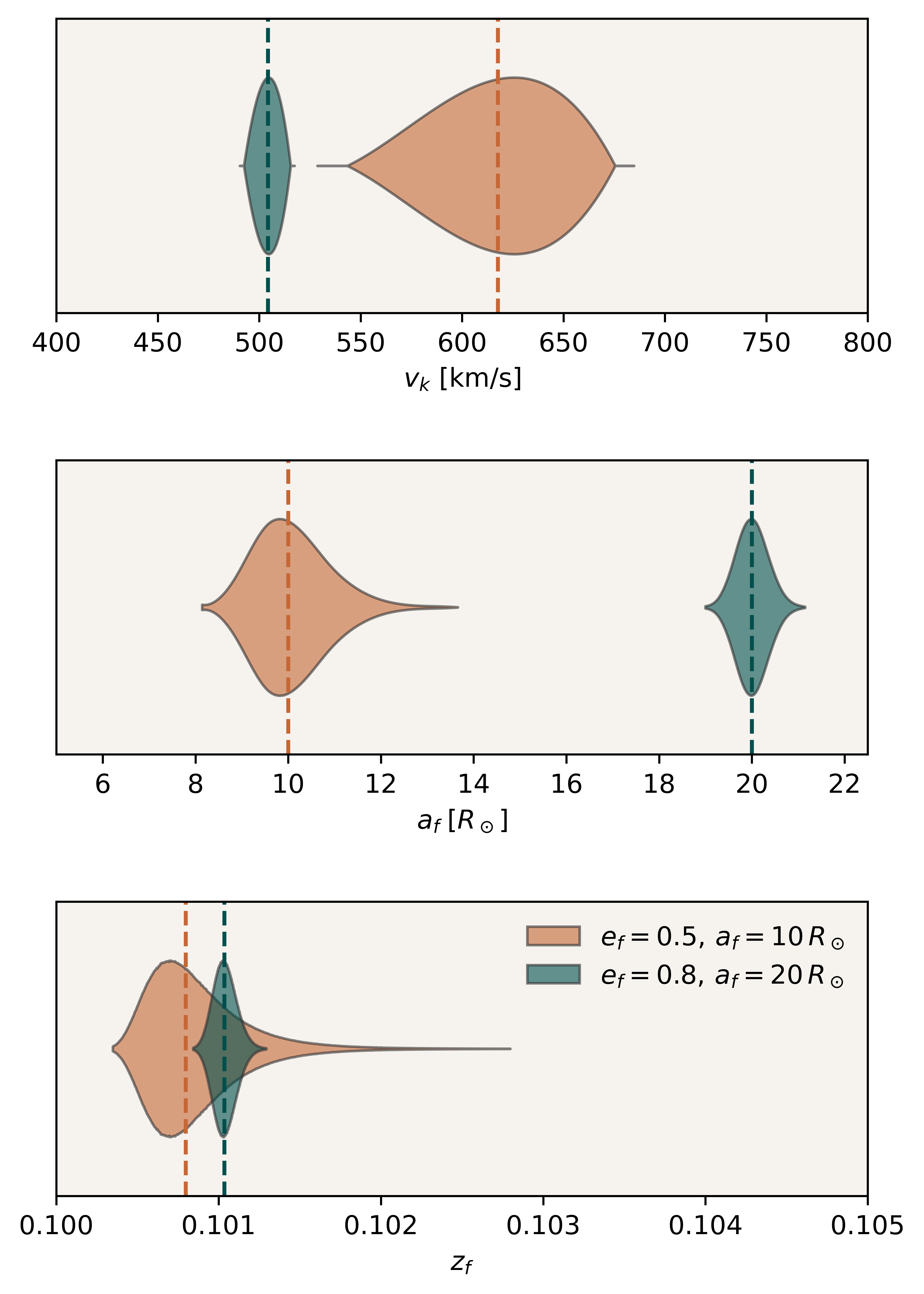}
    \caption{\oldedits{Posterior distributions for the inferred formation parameters of a BBH system, assuming multiband GW observations for two scenarios for a BBH with masses $m_{1} = m_{2} = 30 M_{\odot}$ merging at $z_{m} = 0.1$: one with formation eccentricity $e_{f} = 0.5$ and formation orbital separation $a_{f} = 10 R_\odot$ (orange), and another with $e_{f} = 0.8$ and  $a_{f} = 20 R_\odot$ (teal). The violin plots show the probability distributions for the natal kick velocity ($v_{k}$), the orbital separation at formation ($a_{f}$), and the formation redshift ($z_{f}$). The dashed vertical lines represent the true values for each parameter.} 
}
    \label{fig:multiband}
\end{figure}

\oldedits{
\section{Understanding BBH Formation with Multiband Observations}
\label{sec:multiband}

Our current method infers a BBH’s formation eccentricity ($e_f$) from its spin-orbit misalignment but cannot independently determine properties like the natal kick ($v_k$), orbital separation at formation ($a_f$), or formation redshift ($z_f$). This limitation arises from the degeneracy between the natal kick and orbital velocity: we can measure their ratio ($v_k/v_\text{orb}$) from the spin tilt, but not the individual values.
Fortunately, multiband gravitational-wave astronomy can solve this problem by combining measurements from ground-based detectors with those from the future space-based LISA observatory. While XG ground-based detectors will measure spin-tilt with high precision, they detect systems that have already had their eccentricity largely erased by gravitational radiation. The Laser Interferometer Space Antenna (LISA), on the other hand, is sensitive to BBH inspirals at much lower frequencies, which means it will detect a residual eccentricity ($e_{\text{LISA}}$) much earlier in their inspiral.

The beauty of multiband observations is that we can use these two different eccentricity measurements---$e_{f}$ from an XG detector's spin-tilt and an independent measurement of the residual eccentricity ($e_{\text{LISA}}$) from LISA~\cite{Nishizawa:2016jji,Buscicchio:2021dph}---to break the degeneracy in the formation parameters. The rate at which eccentricity decreases is directly related to the orbital separation and the component masses~\cite{Peters:1963ux}. With two eccentricity values at different points in the inspiral, we can solve for the initial orbital separation at formation, $a_{f}$. Knowing $a_{f}$ then allows us to calculate the natal kick velocity, $v_{k}$. The time between formation and merger depends on the $a_f$ and $e_f$; knowing this delay allows us to infer the redshift at which the binary formed, $z_f$.

To illustrate this, let's consider two cases for a BBH system with masses $m_{1} = m_{2} = 30 M_{\odot}$ that merges at a redshift of $z_{m} = 0.1$: \textit{i}), where the binary formed with  $e_{f} = 0.5$ and $a_f= 10 R_\odot$, and \textit{ii}), which formed with $e_{f} = 0.8$ and $a_f= 20 R_\odot$. The two cases reflect different evolutionary channels: small $a_f$ from stable mass transfer, large $a_f$ from unstable mass transfer~\cite{vanSon:2021zpk}. In both cases, we assume negligible mass loss for simplicity.
The formation eccentricity $e_{f}$ can be inferred from the spin-tilt measurement with an XG detector assuming a $1^\circ$ accuracy. For these two cases, the corresponding LISA residual eccentricities at $0.01$ Hz would be $e_{\text{LISA}} = \mathcal{O}(10^{-3})$ with an error $\sim10\%$, based on fits from Ref.~\cite{Nishizawa:2016jji}. We neglect uncertainties in the binary masses from XG detector, as these are expected to be much smaller than the errors in the LISA measurement of eccentricity.

Once we have these measured values for $e_{f}$ and $e_{\text{LISA}}$, we can use the physical relationships that govern binary evolution to determine the orbital separation at formation, $a_{f}$. This knowledge then allows us to calculate the natal kick velocity ($v_{k}$) and the formation redshift ($z_{f}$). In Fig.~\ref{fig:multiband} we show the violin plots for the two cases and we find that for both of the cases explored, the multiband observations allow us to constrain $a_{f}$, $v_{k}$, and $z_{f}$  within a 10\% error margin. Measuring these quantities would allow us to constrain the host environments, mass transfer, common-envelope evolution, and core-collapse dynamics.

}

\section{Conclusions}
\oldedits{
We present a method to infer the orbital eccentricity of a binary black hole at formation from measurements of spin-orbit misalignment. By using the natal kick from the second supernova as a link between spin and orbit, this approach circumvents the difficulty of directly detecting eccentricity and provides a novel way to probe the immediate aftermath of core-collapse supernovae.}

\new{
Applying this method to GW190412 and GW241011 demonstrates that spin-orbit misalignment can be mapped into meaningful constraints on the system's formation eccentricity ($e_f$). While the broad spin-tilt posterior for GW190412 yields only broad constraints, the significantly tighter spin-tilt constraint for GW241011 shows that this indirect method is already a viable tool for precisely inferring $e_f$ for high-quality events. These case studies thus reinforce the potential of the approach as a proof of principle.
}

\oldedits{Looking ahead, advances in detector technology will make such indirect measurements increasingly precise. In particular, multiband gravitational-wave observations using this approach will allow independent determinations of a binary’s formation redshift, birth orbital separation and natal kicks, offering information on the environments and processes that shape these systems. Measuring the formation redshift will help us understand the environment where the binary was born, constraining the metallicity and star formation history of its host galaxy. Determining the orbital separation at birth will provide new constraints on critical yet poorly understood processes such as common-envelope evolution and mass transfer. This method will also allow us to measure natal kicks received by BHs in compact binaries. Unlike wider X-ray binaries, which are easily disrupted by large kicks, the tight orbits of gravitational-wave sources allow them to survive even very strong kicks. By studying this previously inaccessible population, we can explore higher kick magnitudes and gain a more complete understanding of core-collapse dynamics. This method provides a pathway to relate gravitational-wave observations with the immediate consequences of BBH formation.}

\section*{Acknowledgements}
VB thanks Shrobana Ghosh for comments on an earlier draft of this manuscript.
V.B. acknowledges support from the  NASA Hubble Fellowship grant HST-HF2-51548.001-A awarded by the Space Telescope Science Institute, which is operated by the Association of Universities for Research in Astronomy, Inc., for NASA, under contract NAS5-26555.

\bibliography{ref}

\end{document}